# Quantum Efficiency Enhancement by Mie Resonance from GaAs Photocathodes Structured with Surface Nanopillar Arrays

Xincun Peng[1,2,3,4] Zhidong Wang[1], Yun Liu[1], Dennis M. Manos[2], Matt Poelker[3], Marcy Stutzman[3], Bin Tang[1], Shukui Zhang[3,5], Jijun Zou[1,2,3,6]

A new type of negative electron affinity (NEA) photocathode based on GaAs nanopillar-array (NPA) Mie-type resonators was demonstrated for the first time. For visible wavelengths, the Mie resonances in GaAs NPA reduced light reflectivity to less than 6% compared to a typical value > 35%. Other benefits of NPA resonators include an enhanced density of optical states due to increased light concentration and increased electron emission area. These features resulted in improved photoemission performance at the resonance wavelength demonstrating maximum quantum efficiency 3.5 times greater than a GaAs wafer photocathode without the NPA structure. This optically "dark" photocathode (sub-percentage light reflectance over visible wavelengths) but possessing electrically "high-brightness" (enhanced electron emission) provides new opportunities for practical applications such as large-scale electron accelerators, high-resolution night-vision imaging and low energy electron microscopy.

Optical resonance can be excited by the interaction between free-space propagating light and sub-wavelength nano-structured materials, which has been shown to be highly effective for light management in optoelectronic devices[1,2]. Recently, semiconductor sub-wavelength nanostructures formed in materials with high indices of refraction have attracted increased attention because of their ability to support geometrical resonances (Mie-type, plane-wave scattering from spheres, cylinders, or other shapes) with very low parasitic-power absorption loss, and for their compatibility with semiconductor device processing[3,4,5]. In this study, we investigated Mie-type nanostructured resonators to enhance the photoemission properties of semiconductor photocathodes which we believe could foster new opportunities for generating high quality electron beams using negative electron affinity (*NEA*) GaAs-based or similar semiconductor photocathodes[6,7] for large scale electron accelerators[8,9], advanced light sources[10,11], high-resolution night-vision imaging[12,13], low-energy electron microscopes[14,15] and electron beam lithography[16]. Some of these applications require photocathodes that provide high electron spin polarization, *ESP*, but with photoemission quantum efficiency, *QE*, higher than available today[6,7,8,14]. The highest *ESP* reported to date (92%) was obtained using strained super-lattice *NEA* photocathodes.[17,18] However, these photocathodes cannot meet the increasing demands for both reasonable lifetime and high *QE*[17] for proposed next-generation large scale electron accelerator facilities,[19] where both high brightness and high average-current polarized electron beams are required. Photocathodes with low *QE* would require excessive laser power to generate the desired high current electron beams, and photocathode heating due to the laser power deposited into the cathode substrate leads to *QE* decay and unacceptably short operational lifetime[20,21]. Some new GaAs-based photoemission thin film structures[22,23,24,25,26] have been explored to improve *QE* however, the performance of these film structured devices has been limited by the mismatch between the carrier diffusion length and optical absorption depth[27]. In addition, the high index of refraction of GaAs causes up to 35% of the incident light to be reflected from the surface. Scattered laser light at the edge of the photocathode can generate unwanted photoemission that strikes the vacuum chamber walls degrading the vacuum and reducing photogun operating lifetime [28].

---

[1] Engineering Research Center of New Energy Technology of Jiangxi Province, East China University of Technology, Nanchang, 330013, China.
[2] The College of William and Mary, Williamsburg, Virginia, USA.
[3] Thomas Jefferson National Accelerator facility, 12000 Jefferson Avenue, Newport News, Virginia, 23606, USA.

[4] xcpeng@ecit.cn, [5] shukui@jlab.org, [6] jjzou@ecit.cn

A Mie resonator is a nanostructured device that confines and concentrates the incident light in a well-defined spatial mode profile inside the nanostructure by exciting both electric and magnetic dipoles and higher-order multipole resonances. A strong Mie resonance effect is often excited in materials with high index of refraction[1]. In the visible spectral range, recent work has focused on Si semiconductor nano-resonators with large scattering cross sections, to couple and confine light in optoelectronic applications[29, 30, 31, 32, 33, 34, 35, 36]. To date, however, little attention has been paid to GaAs Mie resonators for photocathode applications. Compared to the low visible-light absorptivity associated with the indirect-bandgap of Si, nano-resonators in direct-bandgap GaAs promise much higher natural absorption. Also, in nanostructured GaAs the process of light absorption can be decoupled from carrier collection by strong light concentration[37, 38]. This has been demonstrated in GaAs nanowire resonator solar cells, where light absorption enhancement has been increased by a factor of ~70 with respect to flat films, yielding efficiencies beyond the Shockley-Queisser limit[38]. But despite this exciting result and others [39, 40, 41, 42], it remains very difficult to fabricate p-n junctions and electrodes of typical diode-type device applications such as solar cells using nano-scaled resonators. In contrast, for a GaAs *NEA* photocathode, where the required bias can be applied externally[13] only *p*-doped GaAs is needed, so the fabrication process is much simpler.

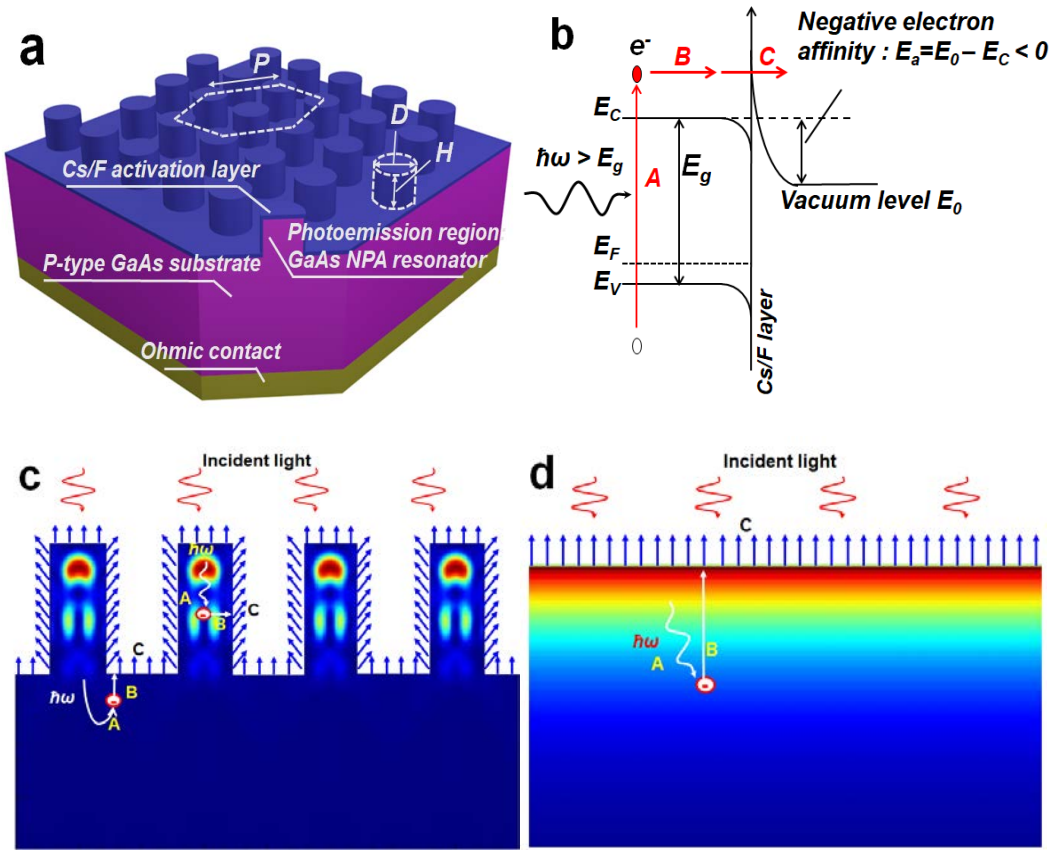

*Figure 1. Hexagonal GaAs NPA resonator photocathode. **a**, The basic device structure of the GaAs NPA photocathode. **b**, The electronic band structure and photoemission process of the NEA photocathode based on Spicer's model, with letters A, B and C signifying the process of photoelectron excitation, electron transportation to surface, and escape from the surface into vacuum, respectively. **c, d,** The photoemission processes of GaAs NPA and flat wafer photocathodes, respectively. The red-color regions denote areas where light absorption in greatest and the blue arrows indicated the trajectories of photo-emitted electrons.*

The basic structure of the *NEA* photocathode with nanostructured GaAs Mie resonators is shown in Fig. 1a. The surface of the photocathode is composed of a hexagonal nanopillar-array (*NPA*) fabricated on *p*-doped GaAs substrate that can be activated to the *NEA* state by co-adsorption of Cs and F [20]. The geometric parameters of the *NPA* include the period of the hexagonal lattice *P*, pillar diameter *D*, and pillar height *H*. Photoemission can be described using Spicer's three-step model[7] comprised of photoelectron excitation, electron transportation to surface, and escape from the surface into vacuum, as illustrated in Fig. 1b, which shows a simplified version of the electronic band structure for the *NEA* photocathode. These steps are illustrated in Fig. 1c for the *NPA* photocathode, to compare to a flat-surface GaAs photocathode referred to hereafter as a *flat wafer* photocathode, in Fig. 1d.

Extending Spicer's model to permit modeling of the *NPA* photocathode[43], quantum efficiency (*QE*) was calculated by considering the probability of absorbing an incident photon to generate an electron in the conduction band ($P_g$), the probability of the electron being transported to the surface of the material ($P_t$), and the probability of an electron at the surface being emitted into the vacuum ($P_e$). Photoemission *QE* is then given by,

$$QE(\lambda) = P_g(\lambda) \cdot P_t(\lambda) \cdot P_e(\lambda) \quad (1)$$

From Fig. 1, we can see the advantages of *NPA* photocathode over the simple *flat wafer* photocathode in terms of photoemission performance. The *NPA* photocathode has a larger effective electron emission area ($a_{NPA}$) compared to a flat wafer ($a_{flat}$) due to the fact that the electrons are emitted from both the top and side surfaces of nanopillars, resulting in larger $P_e$ and therefore, larger *QE*. For the hexagonal *NPA* structure in our study, the ratio of $a_{NPA}$ to $a_{flat}$ ($r_{emi}$) provides a measure of the geometric surface area enhancement and can be calculated by,

$$r_{emi} = a_{NPA}/a_{flat} = \frac{\pi DH + \sqrt{3}P^2}{\sqrt{3}P^2} \quad (2)$$

Furthermore, light absorption in the *NPA* photocathode is larger because the Mie resonances concentrate light within a localized area. This is illustrated by considering the light concentration factor, *C*, defined as the ratio of the absorption cross-section ($\sigma_{abs}$) and the projected physical area ($a_{phy}$) of the *NPA*. For a *NPA* photocathode with surface area *a*, the concentration factor *C* can be written as[44]

$$C(\lambda) = \frac{\sigma_{abs}(\lambda)}{a_{phy}} = \frac{\eta_a(\lambda) \cdot a}{f \cdot a} = \frac{\eta_a(\lambda)}{f} \quad (3)$$

where $\eta_a$ is the light absorption efficiency in the *NPA* and *f* is the array filling fraction. For the hexagonal *NPA* with pillars oriented vertically on the substrate and with the light incident along the axis of the nanopillar, *f* can be calculated as

$$f = \frac{\pi}{2\sqrt{3}} \cdot \left(\frac{D}{P}\right)^2 \quad (4)$$

Strongly enhanced light concentration at the resonance wavelength means photoelectrons can be highly localized inside the nanopillars close to the pillar surface where electrons can be efficiently transported and emitted to vacuum, which serves to enhance $P_g$ and $P_t$, and ultimately leads to enhancement of *QE*.

To understand the optical properties of the NPA Mie-type resonator, a finite-difference time-domain (FDTD, Lumerical FDTD Solutions)[45] analysis was used to model the light absorption spectrum, $\eta_a$, and light concentration factor, *C*, of the GaAs hexagonal *NPA*s as a function of the geometric parameters defined above and the wavelength, $\lambda$, of the plane-wave radiation propagating along the nanopillar axes (normal to the plane of the substrate). The results shown in Fig. 2 illustrate how the geometric properties of the *NPA* structure influence $\eta_a$ and *C* within the photocathode. Figure 2a explores the dependence of $\eta_a$ on pillar diameter, *D*, for pillar height *H* = 750 nm and lattice period *P* = 600 nm. The *NPA* filling fractions, *f*, are also shown in the figures and were used to calculate the light concentration coefficient *C*. Some dominant absorption spectra branches are present in Figure 2, similar to those reported using nanostructures of Ge[34] and Si[35, 36]. These absorption branches result from Mie resonances and can be explained with Mie's scattering theory[1], namely, the resonant-orders are determined by the ratio of *D* to $\lambda/n$ with *n* being the refractive index. Specifically, the lowest-order magnetic/electric dipole (*MD*/*ED*) mode is excited when ($Dn$)/$\lambda$=1, whereas the quadrupole (*MQ*/*EQ*), and higher-order multipole modes, are excited for larger values of (Dn)/λ which are labeled at the corresponding $\eta_a$ peaks in Fig. 2a. Even though the resonance wavelengths and modes are sensitive to D, the values of the resonance enhanced $\eta_a$ for higher order modes do not vary much with diameter above D = λ/2. It can be seen in Fig. 2a that all of the modes show similar resonance enhanced $\eta_a$ however, the dipole modes have much smaller D and f, and therefore provide a much larger C.

Our simulations demonstrate that the dipole and higher-order modes exhibit similar $\eta_a$ dependence on H and P NPA resonators. To simplify the discussion, we present results of dipole-dominated resonators here. Figure 2b illustrates the dependence of the light absorption spectra $\eta_a$ on H for the dipole mode NPA resonators with D = 120 nm and P = 600 nm. Similar to Si[36], larger pillar height leads to stronger resonance electromagnetic field coupling in the nanopillars and provides stronger absorption peaks. Figures 2c and d explore the dependence of $\eta_a$ spectra on the pillar lattice period for dipole mode NPA resonators with pillar diameter 120 nm and pillar heights H = 350 and 750nm, respectively. Increasing P decreases f and also decreases the field coupling in nanopillars which significantly weakens the $\eta_a$ peaks. However, larger field coupling can be achieved using taller nanopillars which allow larger P for the highest $\eta_a$ peak and also larger light concentration C due to smaller f. With respect to the dipole (D = 120 nm) and the quadrupole (D = 260 nm) modes in an NPA with pillar height 750 nm and period 600 nm, the highest $\eta_a$ values were ~91% and ~92% near the resonance wavelengths 650 and 610 nm, respectively. The corresponding light concentration values C were ~25 and ~5, respectively.

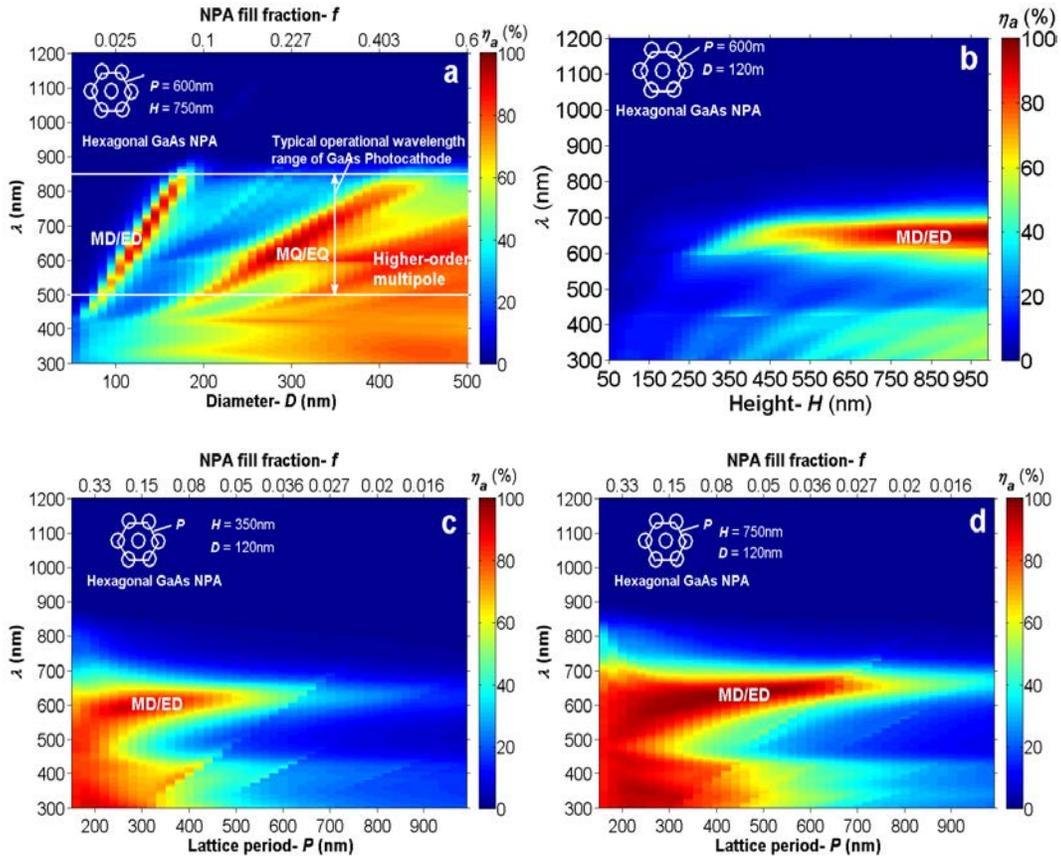

*Figure 2. Simulated light absorption spectra in GaAs NPA resonators. Absorption percentage indicated by color scale in all figures: **a**, Dependence of $\eta_a$ (color) spectra upon diameter D when P = 600 nm and H = 750 nm. **b**, Dependence of $\eta_a$ (color) spectra upon height H for dipole NPA resonator when P = 600 nm and D = 120 nm. **c, d,** Dependence of $\eta_a$ (color) spectra upon lattice period P for dipole NPA resonator when H = 350 nm and 750 nm, respectively.*

Based on the various dependencies illustrated by the plots shown in Fig. 2, a design philosophy was followed to achieve the largest QE enhancement for the NPA photocathodes. Namely, for a practical NPA photocathode operating at a specified wavelength, the nanopillar diameter, D, was selected to excite the corresponding resonance mode, with the nanopillar height, $H$, large enough to fully absorb the incident light and the lattice period, $P$, selected to balance the light concentration and the electron emission area of the nanopillars. From Fig. 2a, both the dipole and quadrupole resonance modes dominated the $\eta_a$ absorption spectra branches covering the wavelength range 500-800 nm, consistent with photoemission requirements of GaAs photocathodes. The largest $C$, and potentially the strongest $QE$ enhancement, can be achieved using the MD/ED resonance mode. Unfortunately, it was difficult to fabricate the GaAs *NPA* photocathodes with sufficiently small $D$ (<150nm) to excite the MD/ED resonance over the entire 500-800 nm wavelength range while simultaneously providing large enough pillar height $H$ (>700nm) to fully couple and absorb the incident light. As a result, this work focused on $QE$ enhancement within the wavelength range 600-700 nm.

One GaAs NPA photocathode sample was fabricated with a dipole resonance mode ($P$ = 600 nm, $H$ = 350 nm, $D$ = 120 nm, labeled as sample *S1*) and two samples were fabricated with quadrupole resonance modes ($P$ = 600 nm, $H$ = 750 nm, $D$ = 260 nm and 280 nm, labeled as samples *S2* and *S3*). The samples were fabricated using substrate-conformal imprint lithography[45] (SCIL) and inductively coupled plasma[46] (ICP) etching processes. A detailed description of the fabrication process is provided in the appendix. An SEM image of a fabricated GaAs *NPA* is shown in Fig. 3a. Surface reflectance spectra for these three *NPA* samples together with reflectance measurements of a flat wafer GaAs sample (labeled as W1) are shown in Fig. 3b. It can be seen that the measured reflectance spectra and resonance peak wavelengths are in good agreement with calculations. Measurements indicate the reflectance of flat wafer sample was greater than 35% across the entire visible spectral range, whereas the lowest reflectance of all of the *NPA* samples was less than 6% at wavelengths between 600~700 nm. It should be noted that light absorption by the pillars of sample *S1* will be smaller than that of samples *S2* and *S3* because sample *S1* has much smaller pillar height. In this sample (*S1*), unabsorbed light will be transmitted and absorbed by the substrate. Optical reflectivity measurements cannot distinguish exactly where the light is absorbed but the photoemission *QE* measurements presented below provide some information.

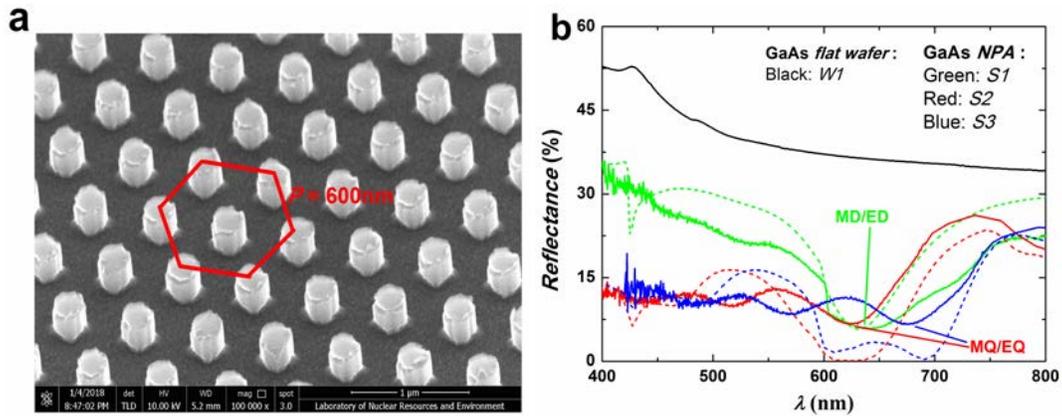

*Figure 3. Surface morphology and reflectance spectra of GaAs NPA fabricated by SCIL on GaAs substrate. a, SEM image of the NPA on GaAs substrate. b, Measured (solid line) and calculated (dash line) reflectance of GaAs NPA samples S1 (D =120nm, H = 350nm, P = 600nm), S2 (D =260nm, H = 750nm, P = 600nm) and S3 (D =280nm, H = 750nm, P = 600nm). The reflectance of a GaAs flat wafer sample W1 was also measured and shown in b.*

The GaAs *NPA* and *flat wafer* samples were successfully activated to an NEA state inside an ultrahigh vacuum chamber with base pressure of ~$10^{-11}$ Torr (see appendix for a description of the photocathode activation process). Measured QE spectral response curves for all GaAs *NPA* samples are presented in Fig. 4a. Resonance modes can be seen in the *QE* spectra, and wavelength positions of *QE* maxima are in agreement with Mie resonance mode for each corresponding geometry predicted from the computational model. Figure 4a shows that across the entire visible spectrum all of the modes associated with all of the *NPA* samples had *QE* values greater than the *flat wafer* sample *W1*. Although the damage-free, "epi-ready" flat samples, W2, had higher QE than dipole sample, the values for that sample were generally lower across the visible than the taller quadrupole sample. At peak resonance wavelengths (λ = 620 nm for *S1*, λ = 627 nm for *S2*, and λ= 655 nm for *S3*), the *NPA* samples exhibit *QE* enhancement by factors of 1.8, 3.5 and 3, respectively.

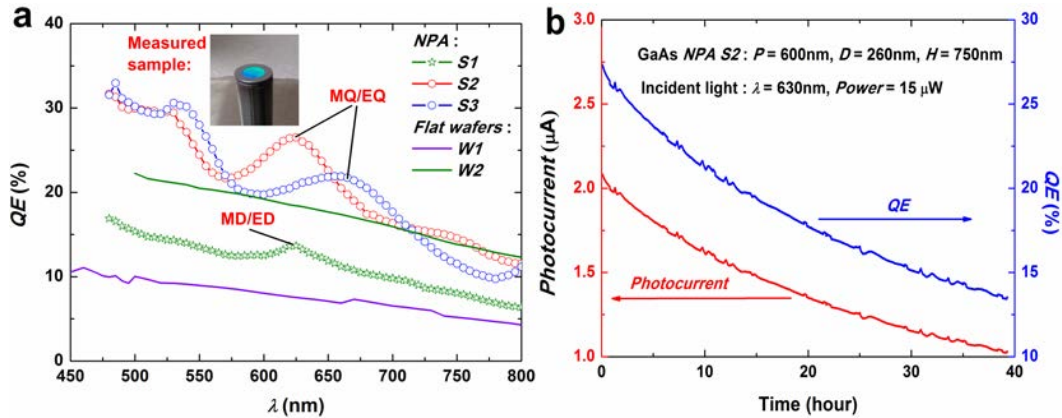

*Figure 4. Measured QE of the GaAs NPA and flat wafer photocathode.* **a,** *QE spectra of the GaAs NPA samples S1 (D =120nm, H = 350nm, P = 600nm), S2 (D =260nm, H = 750nm, P = 600nm), S3 (D =280nm, H = 750nm, P = 600nm) and GaAs flat wafer samples W1(flat wafer exposed to air in this work), W2(Epi–ready flat wafer reported previously[20]) together with a photograph of one of the photocathode samples mounted to a holder for installation in the ultrahigh vacuum test chamber.* **b,** *Dependence of photocurrent and QE on operational time for GaAs NPA sample S2 under the incident light wavelength of 630nm and power of 15μW.*

It should be noted that all of the photocathode samples described in this report were directly exposed to air for several months before testing, and the samples were not chemically cleaned in any manner before installation inside the ultrahigh vacuum chamber. It is well known that GaAs photocathodes are very sensitive to surface conditions[12]. Impurities and defects can be introduced onto the surface during air exposure, sample handling, vacuum chamber bake out, and during the *NPA* etching processes. Surface contamination will increase the electron recombination rate and adversely impact the qualities of *NEA* surface condition and leading to lower *QE*. To gauge the impact of surface contamination, *NPA* photocathode *QE* results were also compared to the best *QE* results of a previously tested epi-ready GaAs *flat wafer*[20] labeled as *W2* and shown as green line in Fig. 4a, which underwent similar handling procedures and the same vacuum chamber bakeout, thermal cleaning and *QE* measurement processes using the same vacuum apparatus. It can be seen that *flat wafer* sample *W1* suffered more surface contamination compared to *flat wafer* sample *W2*, providing only half the *QE*. But even when comparing the *NPA* results to the flat wafer sample *W2*, *QE* enhancement factors of 1.5 and 1.3 at the resonance peaks were still achieved from *NPA* samples *S2* and *S3*, respectively.

The *QE* stability (also referred to as *QE* photocurrent lifetime) of the *NEA* state for GaAs *NPA* sample *S2* was measured (Fig. 4b). When illuminated with 15μW of light at 630 nm, the *QE* of sample *S2* decreased by 50% after 40 hours of continuous photoemission at current > 1μA. This result is quite good compared to measurements performed with similar vacuum chambers. The *QE* stability of a GaAs photocathode is largely dependent on the vacuum (and the bias voltage in case of a high voltage DC gun) conditions under which the measurement occurs. Practical high voltage photoguns used at modern accelerators[8, 10, 19] can be expected to provide better vacuum and therefore better *QE* stability than reported here. The excellent *NEA* activation performance of GaAs *NPA* photocathodes can be also attributed to the larger effective active surface area compared to a *flat wafer* GaAs photocathode. And it is expected that even higher *QE* and better *NEA* stability could be achieved by optimizing the fabrication and surface cleaning processes.

As a final assessment of the GaAs *NPA* photocathode performance, consider once again Spicer's three-step model as described by Eq. (1). The probability of photoelectron generation in the conduction band, $P_g$, was calculated using the FDTD simulations[45]. The photoelectron transport probability, $P_t$, was simulated using the Cogenda technology computer-aided design (TCAD)[23] tools (for detailed description, see Appendix). The measured *QE* data was used to normalize the escape probability, $P_e$, for the electrons. This tunneling process is related to the surface dipoles formed during NEA activation conditions, involving the Cs/F doping. Determining the details of composition, morphology, and crystalline state for these surface species cannot be easily done for NPA, this factor was not computer-simulated. Figures 5a and b show the modeled *QE* spectra for the geometries of samples *S1* and *S2*, as well as the measured *QE*. The *QE* contributions from the *NPA* and the substrate were calculated separately, and illustrate the significant *QE* enhancement from the *NPA*s at the resonance wavelengths. The sum of the two calculated components agrees reasonably well with the measured *QE*, especially for the dipole case. In the quadrupole case, more significant deviations occur in specific regions over the spectral range. Near the resonance region from 575 nm to 655 nm, the total measured value appears to agree better with the quadrupole contribution, possibly indicating little of the expected substrate contribution.

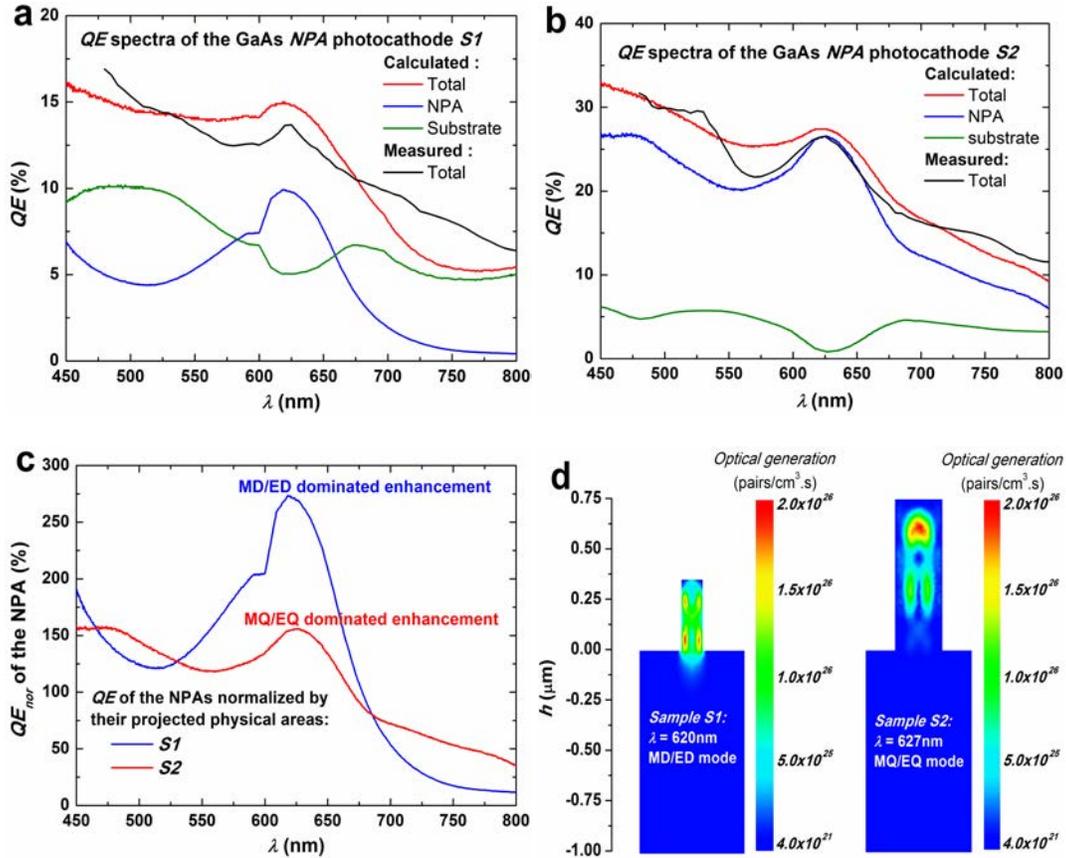

*Figure 5. Analysis of the GaAs NPA photocathodes S1 (D =120nm, H = 350nm, P = 600nm) and S2 (D =260nm, H = 750nm, P = 600nm) by fitting the Spicer's model to the measured QE. a, b, The QE spectra of the samples S1 and S2. c, QE normalized by the projected physical area of NPA ($QE_{nor}$) for samples S1 and S2. d, Distribution of photoelectron generation rate in pairs/cm$^3$·s within the vertical cross section of the samples S1 and S2 in the plane parallel to the E-field of the linearly polarized incident light (light intensity is 1 W/cm$^2$) at the resonance wavelength.*

To better understand the impact of enhanced light concentration, the quantum efficiency of the *NPA* was normalized by its projected physical area ($QE_{nor}$) as reported in other works[35, 39, 40, 41, 42]. Specifically, $QE_{nor}$ was calculated by considering only the incident photons projected onto the physical size of nano-resonator. For the samples *S1* and *S2*, the NPA exhibits $QE_{nor}$ up to 270% and 155% at the quadrupole and dipolar resonance wavelengths, respectively, which further confirms the strong light concentration ability of the *NPA* resonator. Note, $QE_{nor}$ values can exceed 100% because the actual absorbed light power is greater than the computed (projected) light power within the nanopillars[39], so values above 100% do not contradict the definition of *QE*, which assumes a maximum generation of one electron per absorbed photon. A larger $QE_{nor}$ in *S1* than *S2* demonstrates stronger light concentrating ability for the dipole mode and is consistent with the previous analyses. Fig. 5d shows the distribution of the photoelectron generation rate in *S1* and *S2* at their resonance wavelengths. Photoelectrons are mainly generated in the nanopillars, indicating that the density of states is enhanced by the electric fields in the pillars compared to those of the substrate.

Table 1 shows calculated values of $P_g$, $P_t$ and $P_e$ described in Eq. (1) along with performance parameters at the resonance wavelengths for the *NPA* samples *S1*, *S2* and for the *flat wafer* sample *W1* are listed in Table 1. It shows that *QE* is primarily limited by $P_e$, which can be significantly affected by surface damage and contamination arising from the etching and handling processes used for photocathode fabrication. For a unit surface area, the $r_{emi}$ values defined by Eq. (2) were also calculated and listed in Table 1. For the GaAs *NPA* sample *S2*, all of the photoemission performance parameters in Table 1 are larger than for the *flat wafer* sample *W1*, which leads to a *QE* enhancement of 3.5. Whereas for *NPA* sample *S1*, the smaller pillar height of the *NPA* leads to a larger portion of the incident light being coupled into the substrate. Since more light is absorbed, but not as many electrons emitted, the *QE* enhancement is only a factor of 1.3.

*Table 1. Photoemission performance parameters of GaAs NPA and flat wafer photocathodes at the resonance wavelengths ($\lambda_r$)*

| Samples | $\lambda_r$ | Materials | $P_g$ (%) | $P_t$ (%) | $P_e$ (%) | $r_{emi}$ | QE(%) |
|---|---|---|---|---|---|---|---|
| *S1* | 620nm | NPA | 49.4 | 98.5 | 20.0 | 1.42 | 9.9 |
| | | substrate | 40.9 | 54.9 | 22.3 | | 5.0 |
| *S2* | 627nm | NPA | 88.4 | 93.2 | 32.2 | 2.97 | 26.5 |
| | | substrate | 5.60 | 71.3 | 22.5 | | 0.9 |
| *W1* | 620nm | wafer | 63.3 | 51.5 | 23.3 | 1.00 | 7.6 |

In conclusion, we have presented evidence that a rational design of photocathodes using materials with high-index of refraction formed into resonant nanostructures may provide superior performance for photocathodes and other high-brightness emission sources. We demonstrated excitation of Mie-type resonances in GaAs *NPA* that result in significant *QE* enhancement. High *QE* is extremely important for dc high voltage photogun applications, particularly for accelerator applications that demand very high beam-current with long operational lifetime. We suggest that such structures can simultaneously provide a non-reflective surface ("dark" over visible wavelengths) that can be tuned for optimal responses in specific wavelength ranges where high-power, linearly-polarized, time-tunable laser sources are readily available. Such optically dark photocathode surfaces minimize scattered light within the photogun, avoiding unwanted photoemission from the edge of the photocathode that can lead to electron leakage that strikes the vacuum chamber walls degrading the vacuum, lowering *QE* and reducing photocathode lifetime via ion back-bombardment[28]. The results reported here, namely ~27% *QE* and ~90% light absorption, demonstrate the feasibility of this approach. We discussed the need for further work to optimize the fabrication processes, and thereby to reduce damage or contamination during etching, handling, activation, and vacuum practice during operation. These steps will lead to further improvement of *QE* and to longer-life for *NEA* cathodes. It seems reasonable that nanostructured Mie-type resonators can be extended to other materials, in particular the strained-superlattice GaAs-based photocathodes to provide high *QE* and higher spin polarization, which will be an interesting study in future.

## Methods

**Theoretical Methods.** Spicer's three-step photoemission model[7, 43], illustrated in Fig. 1b, was used to study GaAs NEA *NPA* photocathodes. The first step is photoexcitation of valence electrons into the conduction band (process *A* in Fig. 1b), which was modeled by calculating the interaction between incident light and GaAs using a finite difference time domain (FDTD)[45] method. The FDTD setup for simulating the optical properties of GaAs *NPA* photocathodes is shown in Fig. 6a, where the Poynting flux through two test (imaginary) surfaces, one at the top of the GaAs *NPA* resonator, and the other at the bottom, is computed. These are used to compute the light reflection and transmission. A right circular cylindrical three-dimensional test surface was drawn around the GaAs nanopillar to determine the spatial distribution of the electric field $E(\lambda, x, y, z)$ of the incident light (at wavelength $\lambda$). Using the square of the amplitude of the computed field, the generation rate of the electrons $G(\lambda, x, y, z)$ by photoexcitation was calculated as follows:

$$G(\lambda, x, y, z) = \frac{\pi \varepsilon'' |E(\lambda,x,y,z)|^2}{h} \quad (5)$$

where $h$ is Planck's constant and $\varepsilon''$ is the imaginary part of the permittivity for GaAs. Following generation, electrons must be transported to the surface, as shown in process *B* of Fig. 1b. This was simulated by solving the electron and hole drift-diffusion equations based on a full Newton's scheme. A Cogenda visual TCAD tool[23] was used for this simulation. Fig. 6b shows the TCAD setup, in which the bottom and the top surfaces were set as an ohmic contact and charge emission boundary conditions, respectively. The electron transport current $I_t(\lambda)$ was calculated by importing $G(\lambda, x, y, z)$ into the TCAD model. The third step is the emission of electrons into vacuum (process *C* in Fig. 1b). The layer of Cs and F applied to the surface causes the vacuum energy level to drop below the bulk conduction band minima (i.e., the NEA condition). Electrons reaching the surface can tunnel through a narrow surface barrier and be emitted into vacuum. A figure of merit for this process is the surface-electron escape probability $P_e(\lambda)$, which can be obtained by fitting the calculated *QE* to the measured data of the GaAs photocathodes[13].

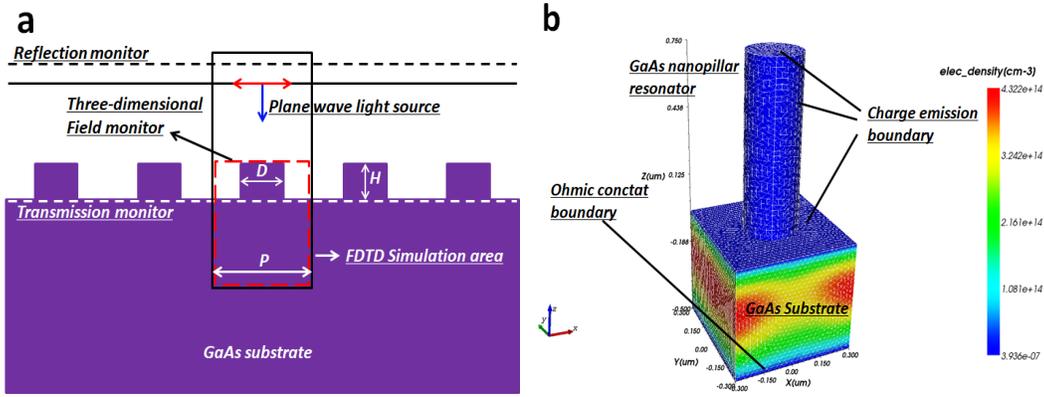

*Figure 6. Theoretical simulation setup of the GaAs NPA photocathode. **a**, Cross-section of the FDTD setup for simulating the optical properties of the GaAs NPA resonators on GaAs substrate. **b**, TCAD setup for analyzing the photoelectric properties of the GaAs NPA photocathode.*

By combining these simulations, the photo-emitted electron current $I_p(\lambda)$ from the GaAs *NEA* photocathode can be calculated as

$$I_p(\lambda) = I_t(\lambda) \cdot P_e(\lambda) \quad (6)$$

Assuming $P(\lambda)$ is the incident light power as a function of wavelength $\lambda$, the quantum efficiency $QE(\lambda)$ of a GaAs *NEA* photocathode will be

$$QE(\lambda) = \left| \frac{h \cdot c \cdot I_p(\lambda)}{q \cdot P(\lambda) \cdot \lambda} \right| \cdot 100\% \quad (7)$$

where $h$ is Planck's constant, $c$ the speed of light, and $q$ the elementary charge.

**Experimental Methods.** The hexagonal GaAs *NPA* was directly fabricated on the p-doped GaAs (100) substrate through substrate-conformal imprint lithography (*SCIL*)[45]. Fig. 7 depicts the fabrication process. A nanopillar array master pattern was first fabricated on a silicon (Si) wafer by electron beam lithography (*EBL*) and reactive ion etching. This *NPA* was then duplicated by first fabricating a polydimethysiloxane (*PDMS*) nano-hole array stamp, which consists of two layers of PDMS: a high-Young's modulus layer (*H-PDMS*) which maintains the features and a flexible low-Young's modulus layer (*L-PDMS*) which allows conformal contact to the substrate. For the GaAs wafer, the first step in patterning was to deposit a layer of silicon dioxide ($SiO_2$) onto the GaAs (100) substrate using plasma-enhanced chemical vapor deposition (*PECVD*). A layer of poly(methyl methacrylate) (*PMMA*) photoresist layer that was spin coated onto the $SiO_2$ layer on the GaAs substrate, then the flexible stamp was applied using a dedicated *SCIL* imprint mechanical tool. The sample then underwent reactive ion etching (Tegal 903e) to transfer the pattern to the $SiO_2$ hard mask. Finally, an optimized inductively coupled plasma etch (Oxford Plasmalab System 100 ICP180) was used to obtain the GaAs *NPA*. The *PMMA* was removed with acetone, and the $SiO_2$ mask layer was then removed with buffered oxide etchant, leaving only the *NPA* structure fabricated on the GaAs substrate.

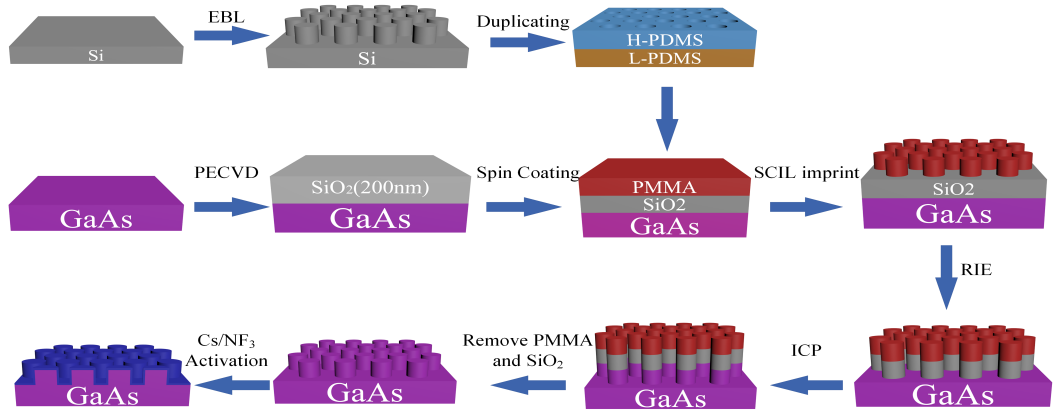

*Figure 7. Schematics of the fabrication process for GaAs NPA NEA photocathodes.*

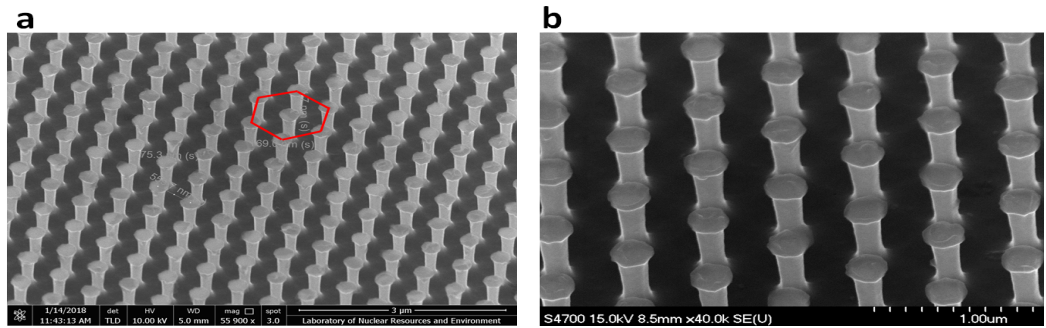

*Figure 8. SEM images of the regular hexagonal GaAs NPA fabricated by SCIL.* **a,** *Before ~600°C heat clean and activation.* **b,** *After 1 hour of ~600°C heat clean and activation.*

The dimensional parameters of the GaAs *NPA* depend on the *SCIL* master pattern and the *ICP* etching time. The morphology of the GaAs *NPA* was imaged with a field emission scanning electron microscope (SEM, NOVA NANOSEM 450). Reflectance spectra were measured using a spectrometer (NOVA-EX, Ideaoptics Instruments, China) at 0° incident angle. Figure 8a shows the *SEM* images of the GaAs *NPA* fabricated by *SCIL*, which show the regular hexagonal *NPA* structures were well aligned and straight.

To obtain the required *NEA* state for the GaAs *NPA* photocathode, the samples were installed in an ultrahigh vacuum chamber with base pressure of ~$10^{-11}$Torr. After installation, the chamber was baked in a hot air oven at 200°C for over 24 hours to remove water vapor. Following the bake, the sample was heated to ~600°C, cooled to room temperature and then activated to reach *NEA* condition using the standard yo-yo activation procedure with cesium and $NF_3$[20]. Figure 8b indicates that the original pattern of the nanowire array remained unchanged, with no deformation or damage after multiple cycles of heat cleaning to ~600°C before chemical activation process. A tunable super-continuum light source (NKT Photonics) provided milli-Watts of well-collimated light from 400 to 850 nm and a picoammeter (Keithley 485) was used to measure the photoemission from the GaAs *NPA* photocathode for the *QE* measurements.

## Acknowledgements

The authors would like to thank Olga Trofimova and Benjamin Kincaid for assistance with the SEM characterizations. This work was partially supported by the National Natural Science Foundation of China (Grant Nos. 11875012, 61204071 and 61661002) and authored by Jefferson Science Associates, LLC under U.S. DOE Contract No. DE-AC05-06OR23177. The U.S. Government retains a non-exclusive, paid-up, irrevocable, world-wide license to publish or reproduce this manuscript for U.S. Government purposes.